Master's thesis

# Network Reconnaissance in IPv6-based Residential Broadband Networks

Tristan Bruns
tristan at tristanbruns dot de

2020-09-23

Betreuung und Erstprüfung
**Dr. Karsten Sohr**

Zweitprüfung
**Prof. Dr. Ute Bormann**

University of Bremen, Germany
Faculty of Mathematics and Computer Science

This page left blank
internationally.

# Contents

# 1. Introduction

Network scanning has been a widely used technique to gather information on the Internet as a whole. Software like ZMap can probe a single TCP or UDP port on the entire public IPv4 address space in under 45 minutes using a gigabit link. [33]

The transition from IPv4 to IPv6 causes traditional network scanning to become less useful. An increasing number of hosts is either IPv6-only or not publicly addressable via IPv4 due to the use of NAT, prompting a need for network scanning techniques for the IPv6-based Internet.

To be able to scan the entire Internet using IPv6, one needs to know the address of every host. For IPv4, this preparative step was simply solved by enumerating all possible addresses. This is no longer viable due to the much increased size (factor $\approx 10^{29}$) of the IPv6 address space, so this step has become a research topic on its own.

All current approaches to IPv6 network scanning make use of lists of IPv6 addresses to be scanned, which are called **hitlists**. A variety of methods for compiling hitlists have been presented, many of which use the domain name system (DNS) or guess addresses based on patterns found in training hitlists. [8, 2, 9] Hitlists created using those methods are useful for scanning server hosts, but do not contain addresses of **client hosts** — smartphones, tablets, PCs, 'smart home' devices, etc. — in a significant amount.

In theory, hitlists containing client host addresses can be created very easily, by collecting client addresses on a server that provides a well-frequented service (e.g. a popular website). [26] This is called a **passive** hitlist source, in contrast to the previously described **active** sources. The obvious issue with passive hitlist sources is that they are usually not publicly available, as they might contain personally identifiable information.

Client hosts are the majority of devices connected to the Internet. Furthermore, when connected to a residential broadband connection, they can send and receive data at substantial speeds, making them attractive targets for botnets. At the same time, in a residential setting, infected hosts are not usually identified due to both lacking know-how and disinterest of the residents.

Providers of residential broadband services provide the customer with *customer-premises equipment* (**CPE**), a single device that provides all the infrastructure needed at the place of residence. It usually contains a modem, but also a router with a stateful firewall and a WLAN access point. While the CPE is connected to the provider's network, the provider routes a small part of its IPv6 address space — the **customer prefix** — to the CPE. This allows the client hosts to use addresses from the customer prefix to communicate over the Internet.

Scanning residential broadband networks is challenging because the IPv6 addresses that take part in the Internet — so called **active addresses** — are changing much more frequently than addresses of server hosts. Client hosts that are mobile (i.e. smart phones) are using temporary addresses — discarded after 24 hours at most — to protect their user's privacy. Furthermore, residential broadband providers often reassign the customer prefixes daily for both privacy and commercial reasons, in a practice called **prefix rotation**. In consequence, passive hitlists containing client host addresses must be expected to be outdated after a day, so for the current hitlist-based IPv6 probing approach to work, one has to move fast.

By aggregating a passive IPv6 hitlist, it is possible to both remove all personally identifiable information, allowing for subsequent publication of the aggregated hitlist, and retain information about approximately where customer prefixes are allocated in a provider's network. Using this information, a simple probing method can efficiently search for both customer prefixes and CPE addresses. This two-step approach will be presented in this thesis.

In the following, I will describe this thesis's structure. In chapter 2, I provide background knowledge about various topics in case you require it.

In chapter 3, I adapt an approach for visualizing the IPv4 address space for use with IPv6 and compare this approach to the current state of the art of visualizing the IPv4 address space, which I argue is worse.

In chapter 4, I describe how to compile IPv6 hitlists containing mostly client hosts and CPEs by taking part in the NTP Pool Project. Additionally, I describe how to create aggregated hitlists to only store information that is useful for longer than a day.

In chapter 5, I present prior work in the field of IPv6 probing, introduce my own simple probing method aimed at finding active customer prefixes and CPEs in residential broadband networks, and present my software implementing this method.

In chapter 6, I test the methods I introduce, by applying them to the three major German residential broadband providers, which have 25 million customers in total. I identify the relevant parts of the three networks by visualizing the aggregated hitlists and use my probing method to search for customer prefixes and CPEs.

Finally, I conclude and provide an outlook in chapter 7.

# 2. Background

This chapter establishes a baseline of knowledge as a foundation for the following chapters.

## 2.1. Internet protocol version 6 (IPv6)

The Internet protocol version 6 (**IPv6** or **IP**) allows **nodes** to exchange packets in the **Internet**. Nodes can either be **hosts**, which only send and receive packets, or **routers**, which additionally forward packets on behalf of other nodes so that the packets can reach their destination.

An **IP address** is an 128 bit unsigned integer used to address nodes in the Internet. The most straightforward method for textual representation of an address — which I will call the *exploded* form — is to display it as a 32 digit hexadecimal number and adding a colon character ':' between all eight groups of four digits each. Usually, an IP address is *compressed* for display, by omitting leading zeros in each group and replacing the longest run of all-zero groups with '::'. In both exploded and compressed form, every hexadecimal digit represents four bits — a **nibble** — of the address. The reverse is true only in exploded form.

A distinction is made between **unicast** and **anycast** addresses: Packets addressed to an unicast address reach the single node with that address. However, an anycast address is used by a group of nodes in addition to their individual unicast addresses. Packets sent to that address reach *any* one of the nodes with that address, hence the name.

The **CIDR notation**[1] appends a forward slash '/' and a **prefix length** from 0 to 128 to an IP address. (e.g. 2001:db8::5/48) Starting with the most significant bit of the address, the given number of bits are called the *prefix* part.

**IP address prefixes** — which are represented in CIDR-notation — are fundamental for both address allocation and routing in the Internet. An IP address is said to be *inside* a prefix if the address starts with the bits of the prefix. The prefix `::/0` has no bits and contains all addresses; with each additional bit the number of addresses in the prefix is halved. A prefix of length 128 contains only one address. It is common to talk about, for example, **a /42 prefix**: This means a prefix with 42 bit prefix length. Linguistic confusion can arise from the fact that a **longer** prefix is a **smaller** prefix (i.e. contains less addresses), and a **shorter** prefix is **larger**. To avoid confusion, a prefix with more bits can be called **more specific** and one with less bits **less specific**.

---

[1] CIDR stands for 'classless inter-domain routing', a routing concept that was introduced for IPv4 in the early nineties. This is completely irrelevant in the current time — except for historical contemplation — and it is fine to just use CIDR as a word. (It is pronounced like 'cider'.) If you insist, read about *classful* addressing in *Computer Networks* by Tanenbaum and Wetherall: [30, §5.6.2]

Table 2.1.: Format of an ICMPv6 error message. *Pos* is the byte offset of each field relative to the start of the ICMPv6 header. *Len* is the size of each field in bytes.

| Pos | Len | Field |
|----:|----:|-------|
| 0 | 1 | Error type (1, 2, 3, 4) |
| 1 | 1 | Code |
| 2 | 2 | Checksum |
| 4 | 4 | Unused / MTU / Pointer |
| 8 | * | Invoking packet |

Address allocation in the Internet is done based on prefix subdivision, for example: The global unicast address space uses the prefix `2000::/3`. The sub-prefix `2a00::/11` was assigned to RIPE-NCC — the regional Internet registry for Europe, the Middle East and parts of Central Asia. RIPE assigned the sub-prefix `2a02:8100::/27` to their member organisation Vodafone Kabel Deutschland. This can continue for several more layers.

While individual organisations can configure the routers in their networks however they want, the border-gateway protocol (**BGP**) has been established as a means to dynamically configure routers to forward packets to other organisation's networks.

## 2.2. Internet control message protocol for IPv6 (ICMPv6)

The Internet control message protocol for IPv6 (**ICMPv6**) is a companion protocol to IPv6 and used to report errors, provide diagnostics, etc. The latest specification was published as RFC 4443. [13]

There are two classes of ICMPv6 messages: **Error messages** and informational messages. Error messages are sent by hosts and routers in response to packets they received but could not process or forward for some reason. They follow the format shown in Table 2.1: The first three fields (type, code, checksum) are common in all ICMPv6 messages, even in informational messages, which are indicated by the most significant bit of the type field being set. The significance of the fourth field depends on the type of error message. Starting at an offset of eight bytes, the message contains the invoking packet, starting with its IPv6 header.[2]

Informational messages do not have a common format (except for the first three fields). **Echo request/response messages** are the two types of informational messages used by the well-known `ping` utility. Their format is shown in Table 2.2. A node responding to an echo request (type field set to 128) only changes the type field to 129 (to indicate an echo response) and recalculates the checksum. No further processing is applied. The identifier and sequence number fields are intended to be used to match a response to a request. The data field is of variable size and intended to be filled with test data to check for communication issues correlating with certain packet sizes or contents. [13, §4.1f]

---
[2]This is not specified clearly in RFC 4443, due to severe editorial issues with that document, see [16].

Table 2.2.: Format of an ICMPv6 echo request/response message. *Pos* is the byte offset of each field relative to the start of the ICMPv6 header. *Len* is the size of each field in bytes.

| Pos | Len | Field |
| --- | --- | --- |
| 0 | 1 | Type: 128 (request) or 129 (response) |
| 1 | 1 | Code: 0 |
| 2 | 2 | Checksum |
| 4 | 2 | Identifier |
| 6 | 2 | Sequence number |
| 8 | * | Data |

## 2.3. Interface identifiers (IIDs) and stateless autoconfiguration

For all currently allocated global unicast addresses, the IPv6 addressing architecture mandates that the first 64 bits signify the subnet prefix and the remaining 64 bits are the interface identifier (IID).[3] [7, 4] In this context, a *subnet* is a group of nodes connected to the same *link*, meaning that some link-layer protocol provides a means of communication between all of them.[4] For subnet nodes to communicate with the Internet, it is necessary for all of them to have unique IP addresses. Since the subnet prefix is the same for all nodes in the subnet, their addresses have to differ in the IIDs.

Unique assignment of IIDs to subnet nodes can be done administratively. **Manual configuration** is the process of manually configuring a subnet node to use a certain subnet prefix and IID, while making sure that the same IID is not configured on more than one node. This approach has the most administrative overhead and is consequently not commonly used. This process can be automated using the *Dynamic Host Configuration Protocol* in *stateful* mode (**stateful DHCPv6**). This requires the operation of a DHCP server on the link, which provides nodes with unique IIDs and keeps a database of IIDs that were assigned.[5] [21]

---

[3] The only exception are inter-router point-to-point links, which benefit from using the two addresses in a /127 prefix. They are of no relevance in the context of this thesis, though.

[4] For example, a few PCs connected to an Ethernet switch can form a subnet, because they can communicate without making use of IP routing, by sending packets on the Ethernet link.

[5] This database is the *state* that makes this mode of DHCPv6 the *stateful* mode. In contrast, *stateless* DHCPv6 is used as part of SLAAC only to provide additional host configuration, e.g. addresses of

Usually, unique assignment of IIDs to subnet nodes is instead done in a stateless, automatic and distributed manner, in a process called *stateless address autoconfiguration* (**SLAAC**). Routers on the link will advertise the subnet prefix. Subnet nodes will then automatically select one or more IIDs for use. Several selection approaches will be explained in the following.

### 2.3.1. Stable interface identifiers

A 1998 approach uses a **modified EUI-64** as IID. For an Ethernet link, this means that the *MAC address* of the Ethernet interface of the subnet node, a 48 bit long *extended unique identifier* (EUI-48), is converted to a 64 bit EUI-64 by inserting the 16 bits `0xFFFE` in the middle. This EUI-64 is then *modified* by inverting the unique/local bit (**u/l bit**, the second least significant bit in the first byte) and used as the IID. Since all Ethernet interfaces on a link need to have unique MAC addresses, IIDs are thus ensured to be unique. [5]

This approach generates IIDs that are not only unique inside a subnet, but globally unique, endangering user privacy. For example, an online advertising company can easily recognise a user's smartphone by the IID, even when it is roaming between different WLANs.

An approach for generating IIDs that are stable only as long as the node does not move to a different subnet was proposed in 2014: **Semantically opaque** IIDs are generated by computing a cryptographic hash function over several inputs, such as the subnet prefix and a permanently stored secret key, and using the output as an IID. Due to the length of the IID and the use of a hash function, the probability of the same IID being generated by two nodes is negligible. [11]

Since 2017, the IETF advises against using IIDs based on modified EUI-64s, and recommends semantically opaque IIDs instead. [12]

### 2.3.2. Temporary interface identifiers

To mitigate the privacy issues resulting from modified EUI-64 IIDs, **privacy extensions**[6] were introduced in 2001. An applying subnet node configures regularly changing temporary IIDs and uses them for outgoing connections.[7] In addition, a stable IID is configured to allow for incoming connections. By default, a new temporary IID is generated daily. [23]

The specification aims at nodes generating temporary IIDs at random, and clearing the u/l bit.[8] The specification proposes a complicated and insecure method of generating the supposedly random temporary IIDs, but allows implementers to use a different approach. A 2015 study by Ullrich and Weippl found that Windows 8 used the bad approach from

---

DNS caching resolvers or NTP servers.

[6] Disconcertingly, the plural noun 'privacy extensions' is used to refer to a single extension.

[7] Actually, the use of the word 'connections' is wrong, since the temporary IIDs are not just used for outbound TCP connections, but all POSIX sockets that are not explicitly bound to an address.

[8] This was a misguided attempt to follow the modified EUI-64 format, as explained in RFC 7136. [4]

the specification, while Linux and Mac OS 10.10 seemed to simply use a pseudorandom number generator. [34] Both approaches generate high-entropy IIDs, so the probability of the same IID being generated by two nodes is negligible.

## 2.4. Internet access providers (IAPs)

An Internet access provider (**IAP**) is an organization that provides their customers with Internet connectivity. IAPs often have both business and private customers. The service provided to private customers is called **residential broadband**, because it is provided at the customers place of residence (home) and the IAPs like to promote the service as being broadband. This thesis is focussed on residential broadband, but IAP services for business customers generally work quite similar in most aspects.

A residential broadband customer is provided with a piece of customer-premises equipment (**CPE**) by their IAP, which combines all network components needed on a typical customer's premises in one physical device: A modem or Ethernet interface to connect to the IAPs wide-area network (**WAN**), a router with firewall, an Ethernet switch for the local-area network (**LAN**), and a wireless LAN (**WLAN**) access point.

When a CPE connects to the IAP's WAN, it initially acquires an IPv6 address for itself using one of the means described in section 2.3. Subsequently, the CPE requests a **customer prefix** from the IAP's routing infrastructure.[9] On the IAP's side, an unused customer prefix is selected from a **customer prefix pool** and routing of traffic destined to this prefix to the CPE is established. The CPE can now create one or more IPv6 subnets for the customer's devices to join, providing them with IPv6 connectivity, too. Some IAPs reassign the customer prefixes daily for both privacy and commercial reasons, in a practice called **prefix rotation**.

An important variable in an IAP's addressing plan is the size of customer prefixes. Since every LAN requires a /64 prefix, the maximum customer prefix length is 64 bits. CPEs will usually support operation of multiple LANs (e.g. a guest network in addition to the LAN for residents) and consequentially need shorter customer prefixes.

The broadband forum — an industry consortium — 'suggests [at] least a /60 for home network or SOHO [small office, home office] environments, with a recommended prefix length of /56.' [17, §4.2] RIPE NCC — the regional Internet registry for Europe, the Middle East and parts of Central Asia — adds: 'It is *strongly discouraged* to assign prefixes longer than /56 unless there are very strong and unsolvable technical reasons for doing this.' (emphasis originally set in bold face) [36]

Assuming an IAP would use /56 customer prefixes, a /40 customer prefix pool would allow for 66 thousand ($2^{16}$) customers. IAPs will often use multiple pools because their network is divided geographically.

Table 2.3.: Market shares and number of BGP-announced IPv4 addresses of residential broadband providers in Germany, as of 2019-06-30. Market share data was gathered by VATM (lobby group of German telecom companies). [35, slide 13] Historical BGP data is from the RIPE Stat API, queried for the AS numbers 3320, 31334, 8881, and 9145. [27, Prefix Count]

| Brand (Company) | Customers % | Customers millions | IPv4 addresses millions |
|---|---|---|---|
| Deutsche Telekom | 39.7 | 13.4 | 34.2 |
| Vodafone Kabel Deutschland | 19.9 | 6.9 | 2.0 |
| 1&1 (United Internet) | 12.4 | 4.3 | 1.4 |
| Unitymedia (Liberty Global) | 10.7 | 3.7 | |
| O$_2$ (Telefónica) | 6.4 | 2.2 | |
| EWE | 1.7 | 0.6 | 0.9 |
| Tele Columbus | 1.7 | 0.6 | |
| M-net | 1.4 | 0.5 | |
| NetCologne | 1.2 | 0.4 | |
| Deutsche Glasfaser | 0.2 | 0.2 | |
| Others | 4.6 | 1.6 | |
| Total | 100.0 | 34.6 | |

## 2.5. German residential broadband market

In preparation for the case study in chapter 6, this section will introduce the German residential broadband market.

Table 2.3 provides an overview of the market shares of the top ten residential broadband providers in Germany. Deutsche Telekom, United Internet and Telefónica provide their service via DSL (former telephone lines on the last mile), while Vodafone and Unitymedia provide their service via DOCSIS (cable television network as the last mile). Vodafone and Unitymedia both operated a monopoly in different parts of the country.

From 2018 to 2019, Unitymedia was bought by Vodafone. At the moment, Vodafone owns the entire German cable television network. This thesis will only look at the network of Vodafone Kabel Deutschland, not at the network of Unitymedia, so Table 2.3 still lists the correct number of customers for the two networks.

IPv4 exhaustion is an issue for both Vodafone and 1&1, as Table 2.3 shows: Both networks use far less IPv4 addresses than they would need to use to provide each customer with an address. This proves that both networks must employ carrier-grade NAT (CGN), so that multiple customers can jointly make use of one IPv4 address. All IAPs in Table 2.3 have rolled out IPv6 connectivity for their residential customers years ago, except for EWE, which still solely uses IPv4.

---

[9] If you are curious: This is done using stateful DHCPv6, see [21, §6.3].

## 2.6. Network time protocol (NTP)

The network time protocol (**NTP**) is widely used to synchronize the system clocks of hosts over the Internet. NTP servers are distinguished by their **stratum**: Stratum 1 servers are synchronized to a local reference clock, such as a atomic clock or a GPS receiver. Servers of stratum 2 and above do not have a reference clock and are synchronized to multiple servers of the next lower stratum. [20, §1]

Several companies operate NTP server clusters to provide time of day information for their products or infrastructure. I know of Microsoft, Apple, Google, and Deutsche Telekom. Furthermore, some (national) metrology institutes allow clock synchronisation with their atomic clocks via NTP, such as the German *Physikalisch-Technische Bundesanstalt* or NIST. They usually operate a few server hosts without intricate redundancy setups or load balancing, so they cannot support large user bases and consequently try to maintain a low profile.

For the general public, the **NTP Pool Project** aims to provide a community-operated NTP server cluster. Volunteers operate NTP servers and add them to the pool's database. The pool's infrastructure monitors all registered servers to ensure that they are reachable and have the correct time. The IPv4 and IPv6 addresses of operational servers are advertised through the pool's DNS infrastructure, which serves the domain `pool.ntp.org` and its subdomains. Users of the pool configure their NTP clients to use those domains to find NTP servers for synchronisation. [24]

Every server added to the pool is sorted into a geographic **zone** depending on the country it is hosted in. (e.g. `de` for Germany) The members of a country zone are also members of one of six matching 'continent' zones.[10] (e.g. `europe` for Europe) All servers are members of the global zone, `@`. DNS lookups of `[zone].pool.ntp.org` return only servers from that zone, while lookups of `pool.ntp.org` will try to guess the requesting user's location based on their IP address and return servers from the matching zone.

To work around software that uses only the first IP address returned by a DNS lookup, all domain names can be prefixed with digits from 0 to 3. (e.g. `1.de.pool.ntp.org`) Queries for all domain names return four randomly selected IPv4 addresses, but for historic reasons only queries for domain names starting with `2.` will also return four randomly selected IPv6 addresses.

---

[10]Africa, Asia, Europe, North America, Oceania, South America

# 3. IPv6 address space mapping using a Z-order curve

Visualisation is an established tool allowing humans to recognise structure in data. This chapter explains how the IPv6 address space can be visualised to reveal patterns in the hierarchy of address prefixes, i.e. to create a topological map.

## 3.1. Problem statement

The map should fulfill the following three requirements:

**Grouping**  Addresses with the same prefix should be grouped. Because address prefixes form a hierarchy, prefixes with a common prefix also need to be grouped.

**Locatability**  Once we visually identified a feature on the map, we need to be able to tell what address prefix this feature is in, in order to investigate it.

**Structure**  The map should enable us to recognise patterns in structured networks, such as multiple subnets laid out in the same way.

## 3.2. Prior work

In 2002, Teoh and coauthors described a visualisation technique for IPv4 address prefixes based on an 'Quad-tree encoding', which is fundamentally the same as the Z-order curve based approach described in this thesis. [32, 31]

In 2006, Randall Munroe, as part of his web comic 'xkcd', published the panel 'Map of the Internet', which contains an overview of the IPv4 address space allocations, this time arranged as a Hilbert curve. [22]

This inspired many authors to visualise the IPv4 address space in this manner, while the prior approach fell into oblivion. Hilbert curves were used, for example, by ISI [3] and CAIDA [6] from 2007 on.

I could not find a comparison of Hilbert and Z-order curves for the purpose of address space mapping. Consequently, I will compare the two in section 3.5.

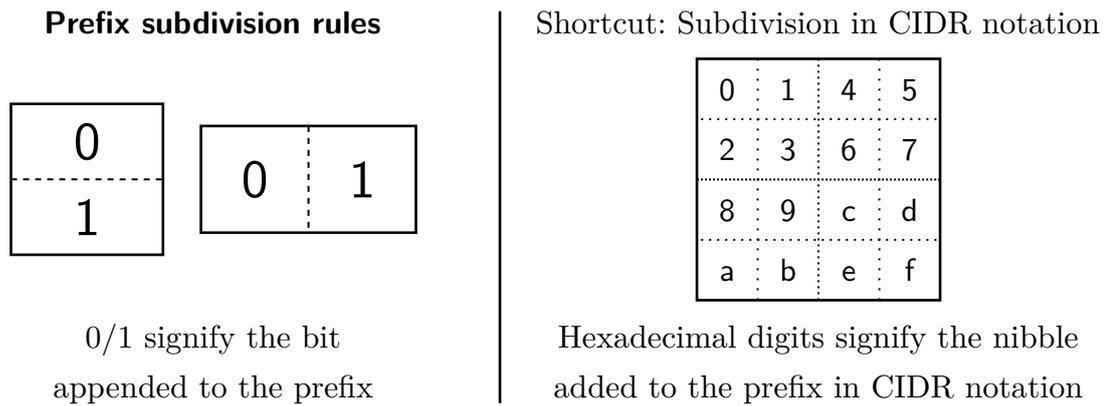

Figure 3.1.: The two prefix subdivision rules plus a shortcut

## 3.3. Mapping using a Z-order curve

Any IPv6 address prefix can be generated by iteratively adding a single bit (0 or 1) to less-specific prefixes, starting with the zero-length prefix ::/0. Correspondingly, the area representing any prefix on the map can be constructed by iteratively splitting areas representing less-specific prefixes according to the following two rules, which are also shown graphically in Figure 3.1:

1. If the prefix has an even length, the square representing it in the map is cut in half vertically. The resulting rectangle on the top/bottom represents the more-specific prefix ending in 0/1, respectively.

2. If the prefix has an odd length, the rectangle representing it in the map is cut in half horizontally. The resulting square on the left/right represents the more-specific prefix ending in 0/1, respectively.

Consequently, as shown in the figure, a shortcut exists for subdividing a nibble-aligned prefix exactly four times, which is equivalent to adding a digit to a prefix in CIDR notation. (e.g. finding 2001:db8:7000::/36 in 2001:db8::/32)

Examples of address space maps are shown in Figure 3.2 and Figure 3.3, for prefixes of even and odd lengths, respectively.

## 3.4. Implementation considerations

The maps display (part of) a $2^{64} \times 2^{64}$ coordinate grid the IPv6 addresses are mapped into using a Z-order curve, as discussed in the next paragraph. This grid has its origin in the top left corner; the X coordinates increase in the right direction and the Y coordinates increase in the bottom direction.

Given a programming environment with 128 bit wide unsigned integers (or bignums), calculating the 2D coordinates according to the Z-order curve is quite straightforward:

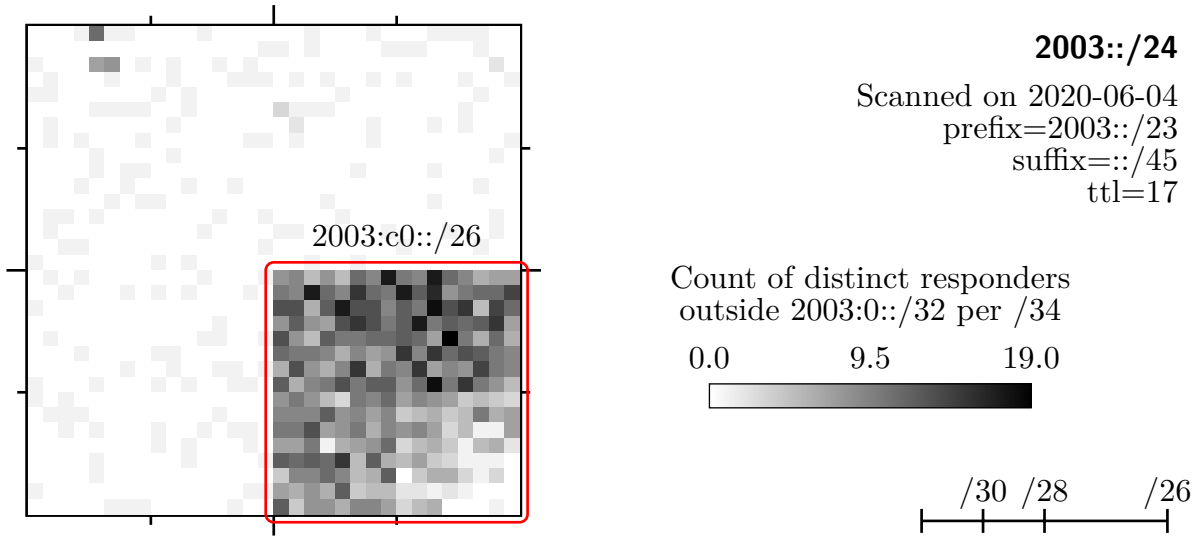

Figure 3.2.: An example of an address space map showing a prefix with even length.

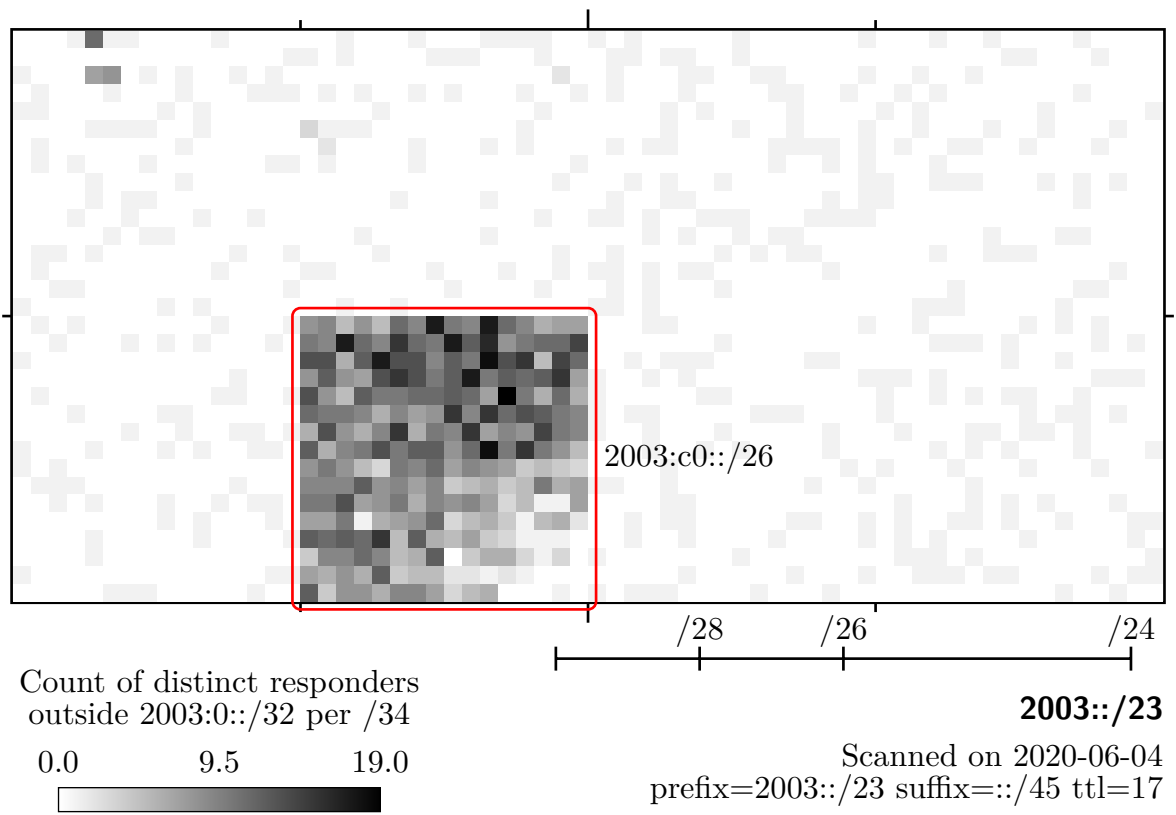

Figure 3.3.: An example of an address space map showing a prefix with odd length.

When an IPv6 address is represented as a number[1] between 0 and $2^{128} - 1$, the X/Y coordinate is the concatenation of the even/odd numbered bits[2] of the address, respectively.

My implementation can be found in the file `ipmap.py` of the source code publication referenced in section A.1.

## 3.5. Conclusion

This section evaluates my Z-order curve based approach for mapping, which adapts Teoh et al. [32, 31] for CIDR and IPv6, and contrasts it with the currently popular Hilbert-curve based approach. See section 3.1 for the list of requirements.

First of all, both approaches group addresses belonging to the same prefix into a rectangle. A closer look at how this grouping is performed is shown in Figure 3.4, which also takes into account the position of the first and last addresses of a prefix. As can be seen, Hilbert curves not only use one additional shape for prefixes of odd length (rectangle in portrait orientation), but they also use four different ways to arrange the addresses inside rectangles of each shape.

Understandably, this makes it very hard to mentally follow the Hilbert curve when reading the map, and therefore obscures where prefixes are displayed. In contrast, my approach supports mental conversion between address prefixes and areas on the map through simple subdivision rules (section 3.3). Furthermore, the map scale allows to judge the size (i.e. prefix length) of features on the map.

As a further consequence of the complicated arrangement of prefixes in the currently popular approach, recognition of structure in networks is hindered, as can be seen in Figure 3.5. In contrast, my approach allows for recognition of the structure.

The differences in arrangement complexity stem from the fact that the Hilbert curve is continuous, while the Z-order curve is discontinuous. In the literature, I could not find any indication as to why using a continuous curve might have advantages and I did not observe any advantages myself. In conclusion, using a Hilbert curve for address space mapping causes severe disadvantages, while offering no apparent benefit.

---

[1] Take care to order the bits and bytes of the address such that they follow the order implied by CIDR. This can be checked by printing the address as a hexadecimal number; it should read the same as the text representation of the address without abbreviation and colons.

[2] Bit numbering starts with 1 at the most significant bit.

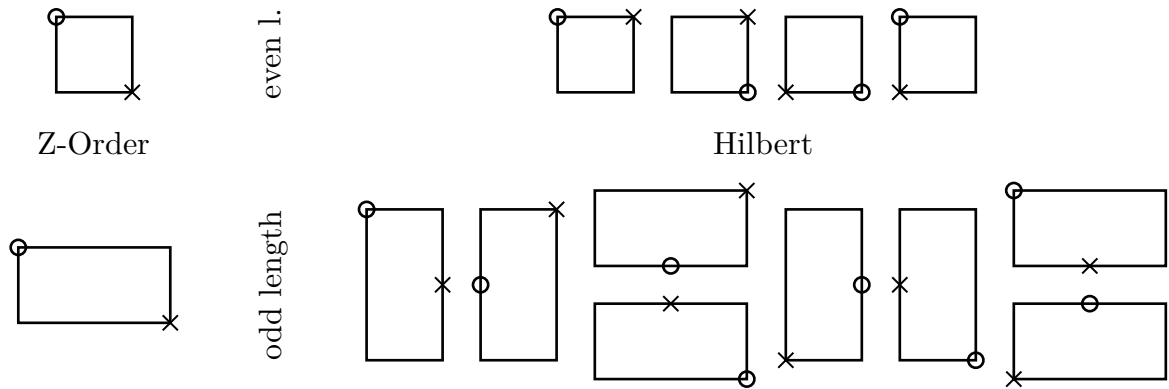

Figure 3.4.: Shapes of an IP address prefix in the map when using a Z-Order or Hilbert curve; for prefixes of odd or even length. The position of the first/last address of each prefix is marked with a circle/cross, respectively.

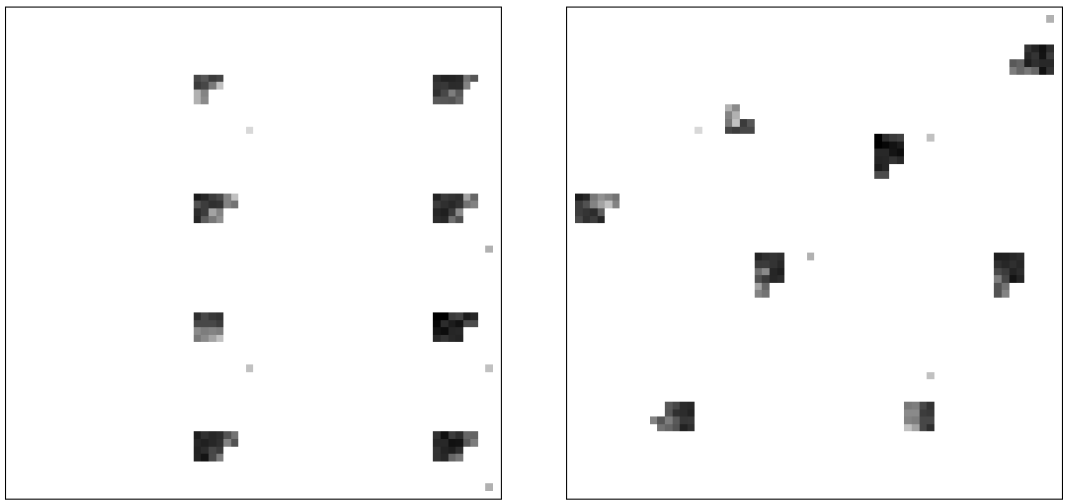

Figure 3.5.: Maps showing measurements of a structured network using a Z-Order/Hilbert curve. (Left/right, respectively) The network consists of eight /37 prefixes, each containing a dynamic prefix pool in the last /40. Each pool contains WAN links in the last /48.

# 4. Hitlist generation through the NTP Pool Project

To facilitate scanning of access provider prefix pools (see chapter 6), they first need to be located. This chapter introduces the NTP pool as a passive, but public, hitlist source that is well suited for this task.

## 4.1. Prior work

Shodan.io — a company selling address list of (vulnerable) Internet services — operated three NTP pool servers until early 2016. They added each server to the pool 15 times with different IP addresses to attract more traffic.[1] Sending any packet to one of the 45 IP addresses triggered an immediate port scan. This was discovered by an NTP pool user in January, 2016. [15] The servers were subsequently removed from the pool by its maintainer. [14]

Plonka and Berger at Akamai — one of the largest global content delivery networks in existence — have analyzed addresses from the network's access logs in their 2015 paper. [26]

## 4.2. Address collection

As a prerequisite, I set up an NTP server and added it to the NTP pool. It was assigned to the zones '@', 'europe' and 'de'. Being a stratum 2 server, it did not require any special timekeeping hardware and could run in a virtual machine. The traffic was in the order of 100 packets per second via IPv6 and 1000 via IPv4.

On the NTP-server host I ran a Python script capturing all the IPv6 packets directed at the host and extracting their source addresses.[2] (`ipcollect` in source code publication, see section A.1) The source addresses were collected in a PostgreSQL database. (schema in `database.psql`)

The collection was started on 2019-11-25 and stopped on 2020-07-19, after running for 237 days. 1219 million packets from 188 million unique source addresses were observed.

---

[1] The pool.ntp.org DNS infrastructure tries to balance the load equally between all IP addresses in each zone of the pool.

[2] I should have narrowed the address collection to packets destined to the NTP server (UDP port 123), but I forgot. Since the host was a single-purpose virtual machine running only the NTP server, I do not expect my mistake to have significant consequences.

Table 4.1.: The top eleven address allocations containing the most source addresses seen at the NTP server. Percentages are relative to the total number of source addresses seen (188 million).

| Organisation | Prefix | Addrs. | % |
|---|---|---:|---:|
| Deutsche Telekom | 2003::/19 | 96 603 113 | 51.3 |
| 1&1 | 2001:16b8::/32 | 18 478 085 | 9.8 |
| Jio (IN) | 2409:4000::/22 | 15 278 900 | 8.1 |
| Bharti Airtel (IN) | 2401:4900::/32 | 14 836 918 | 7.9 |
| Jio (IN) | 2405:200::/29 | 5 773 742 | 3.1 |
| Claro Brasil (BR) | 2804:14c::/31 | 2 394 564 | 1.3 |
| M-net | 2001:a60::/29 | 2 190 965 | 1.2 |
| NetCologne | 2001:4dd0::/29 | 1 816 122 | 1.0 |
| TIM Brasil (BR) | 2804:214::/32 | 1 622 634 | 0.9 |
| Verizon (US) | 2600:1000::/27 | 1 537 359 | 0.8 |
| Vodafone Kabel Dt. | 2a02:8100::/27 | 1 478 030 | 0.8 |

Table 4.2.: Upper section: Quantitative breakdown of the collected IIDs by certain observable features. Lower section: Estimation of how many IIDs were generated using each of the methods from section 2.3.

| | | |
|---|---:|---:|
| IIDs seen | 162 981 920 | 100.0% |
|    …`ff:fe`… | 5 873 844 | |
|    remaining | 157 108 076 | |
|      000… | 10 985 002 | |
|      remaining | 146 129 759 | |
|         u/l bit set | 13 464 131 | |
| Estimated: | | |
|    Manual / stateful DHCPv6 | 10 985 002 | 6.7% |
|    Modified EUI-64 | 5 873 844 | 3.6% |
|    Semantically opaque | 26 928 262 | 16.5% |
|    Privacy extensions | 119 201 497 | 73.0% |

## 4.3. Analysis of address prefixes

Table 4.1 shows the top eleven address space allocations[3] that originated packets to my NTP server. All of them are held by Internet access providers. In preparation for chapter 6, I aggregated the collected addresses by compiling lists of prefixes of a certain length that originated at least one packet:

**hitlist64** contains 134 million /64 prefixes

**hitlist56** contains 44 million /56 prefixes

**hitlist52** contains 8 million /52 prefixes

**hitlist48** contains 1.6 million /48 prefixes

Subsequently, I looked up the prefixes from hitlist48 in the BGP routing table, as published by the University of Oregon Route Views Project, using CAIDA's pfx2as. [28] I found that I had collected addresses from 18891 BGP prefixes announced by 5512 autonomous systems.

## 4.4. Analysis of interface identifiers

The 188 million IP addresses I collected contain 163 million unique interface identifiers (IIDs, see section 2.3). To estimate how many IIDs were generated through each of the means introduced in section 2.3, I split up the set of IIDs using certain heuristics, as explained in the following and also outlined in Table 4.2.

First off, I separated IIDs looking like modified EUI-48s by matching `0xfffe` in the two middle bytes. (I will look into these IIDs later on.) From the remaining 157 million IIDs, I separated those starting with the twelve bits `0x000` and assumed they were configured manually or through stateful DHCPv6.

Going by section 2.3, I expected the remaining 146 million IIDs to have been generated either through privacy extensions or as semantically opaque. Both methods should provide a uniform distribution of IIDs. To check my assumptions up to this point, I plotted a histogram of the values of the first and second byte of all remaining IIDs in Figure 4.1. As you can see, the distribution does in fact look fairly uniform.[4]

The u/l bit allows to estimate the makeup of the remaining IIDs. In semantically opaque IIDs, there is a fifty-fifty chance of the u/l bit being set, while it will never be set by privacy extensions. Consequently, one half of the semantically opaque IIDs can be identified by the u/l bit. The total number of semantically opaque IIDs is estimated to be twice that amount.

---

[3] I am counting per allocation and not per BGP prefix because many access providers announce multiple longer prefixes covering their allocations to have finer control over the traffic entering their network. This practise is called *traffic engineering*. The manner in which an access provider announces its address space can change very suddenly.

[4] There are a few small outliers. Apparently at least one implementation of stateful DHCPv6 can be configured to sequentially allocate IIDs starting not from 0, but from a configurable initial value. In this case, the most visible outliers are caused by sequential IIDs starting with `0xcafe` and `0xabcd`.

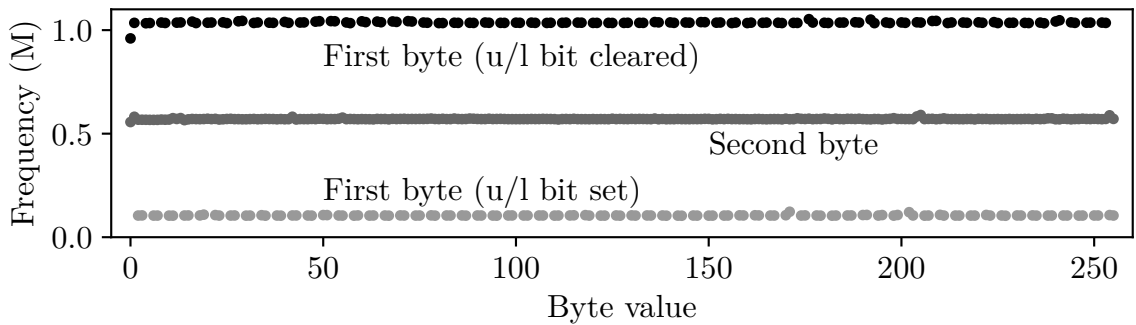

Figure 4.1.: Histogram of the first and second bytes of IIDs I expected to have been generated by privacy extensions or as semantically opaque. The first byte values are shown as two separate data series distinguished by the u/l bit.

Table 4.3.: The top ten vendors from modified EUI-48 IIDs.

| Count | % | Vendor |
|---|---|---|
| 2 385 326 | 41 | Amazon |
| 575 120 | 10 | AVM |
| 456 497 | 7.9 | Arcadyan |
| 417 702 | 7.3 | Samsung Electronics |
| 310 306 | 5.4 | Huawei |
| 251 465 | 4.4 | (unregistered EUI-48) |
| 151 495 | 2.6 | Sonos |
| 150 254 | 2.6 | Sercomm |
| 61 043 | 1.1 | Samsung Electro-Mechanics |
| 45 680 | 0.79 | Motorola |

Finally, I looked up the vendors of the six million modified EUI-48 IIDs. It is important to state that they represent only 3.6% of the collected IIDs and are in no way a representative sample of all IIDs, so they are examined only for additional insight. The top ten vendors are shown in Table 4.3. Amazon[5] and Sonos are vendors of 'Smart Home' devices; AVM, Arcadyan, Huawei and Sercomm manufacture CPEs.

## 4.5. Evaluation and outlook

Running an NTP pool server is a good source for active addresses belonging to client hosts, 'Smart Home' devices and CPEs.

Surprisingly, I found that at least 20% of the users of my NTP server are from outside Europe. (See Table 4.1) This is unfortunate for those users, as time synchronisation is less accurate with increased latency. This could be caused by broken Internet geolocation in the pool.ntp.org DNS infrastructure, export of devices or software aimed at the European market (e.g. defaulting to use `2.europe.pool.ntp.org`) or possibly even widespread use of European DNS resolvers to circumvent censorship. Further research is needed.

---

[5] I verified that this is not caused by virtual machines hosted by Amazon EC2. There are in fact at least 2.4 million *Alexas* in german households, recording their residents' conversations and sending the audio to Amazon, as advertised.

# 5. Probing method

This chapter describes how I conducted each of my network scans and the behaviour of `probe.py` — a scanning software I developed for this purpose.

## 5.1. Prior work

A 2010 paper by Leonard and Loguinov about (IPv4-)Internet-wide service discovery describes the *reverse IP-sequential* scan order I adopted and contrasts it with the well known sequential and uniformly random scan orders. [19]

Beverly and coauthors published a paper about IPv6 topology discovery in 2018. They adapted the IPv4 topology prober *yarrp* to work with IPv6 and released this tool as *yarrp6*. One focus of the paper is the impact of probing load distribution on the effects of ICMPv6 rate limiting. [1]

## 5.2. Rationale for developing my own prober

Yarrp6 is a topology prober, meaning that it sends hop-limited probe packets to prompt every router on the path towards each target IP address to respond. Consequently, many probe packets are sent for each target. Yarrp6 requires a list of target IP addresses to be probed.

My requirements are different: Since I am searching for CPEs, I am really just interested in the last hop, so sending hop-limited probe packets is not required. Furthermore, I want to probe IP addresses that are following certain patterns, since ISP networks are highly regular.

I considered adapting yarrp6 to suit my goals, but upon skimming through its untidy and undocumented code, I decided that it would be easier to write my own untidy and undocumented code. Consequently, I wrote my own prober, `probe.py`, using the programming language Python to lower the bar for reuse.

## 5.3. Target definition

Instead of a list of IP addresses to probe, `probe.py` accepts a list of sets of IP addresses, each called a **range**. A range is defined by a fixed address **prefix** and a fixed address **suffix** and includes all addresses that start and end with the prefix and suffix bit strings, respectively.

Table 5.1.: Structure of an ICMPv6 echo request probe packet. *Pos* is the byte offset of each field relative to the start of the ICMPv6 header. *Len* is the size of each field in bytes.

| Pos | Len | ICMPv6 Field | Value |
|---|---|---|---|
| 0 | 1 | Type | 128 |
| 1 | 1 | Code | 0 |
| 2 | 2 | Checksum | (checksum) |
| 4 | 4 | Ident. + Seq. nr. | Token |
| 8 | 32 | Data | HMAC-SHA-256 over the previous field |

The addresses in each range are not probed in sequential order to reduce measurement artefacts resulting from ICMPv6 rate limiting. [1] As proposed by Leonard and Loguinov, I am using the reverse IP-sequential scan order, which they proved to spread probing load optimally. [19] This order is implemented just like the sequential order, but the bits of the counter are reversed before they are embedded into each IP address. (e.g. a two-bit counter 00, 01, 10, 11 would be embedded as 00, 10, 01, 11)

If the target list contains multiple ranges, they are probed simultaneously. The probe packets are interleaved in such a way that the addresses in each range are probed over the whole runtime of `probe.py`, even if the ranges do not contain the same number of addresses.

## 5.4. Probe packets

Following Beverly and coauthors, I chose to send ICMPv6 probe packets. A single ICMPv6 echo request packet is sent towards each target IP address using a fixed — but configurable — IPv6 hop limit.

The structure of an ICMPv6 probe packet is shown in Table 5.1. The type and code fields are set to 128 and 0, respectively, to indicate an echo request packet. The two 16-bit fields 'Identifier' and 'Sequence Number', which are just echoed back by the responder, are used in combination as a 32-bit field containing the *token*. The token encodes the position of the current target IP address in the target list.[1] The data field of the ICMPv6 echo request packet is used to transmit a message authentication code *(MAC)* computed over the four bytes of the token field. (HMAC-SHA-256 [18, 25])

---

[1] The first n bit of the token are the index of the range in the target list; the remaining bits are the index of the IP address in the range. The value of n is chosen at startup to fit the length of the target list.

Table 5.2.: Structure of an ICMPv6 error message response packet and how it will be parsed by `probe.py`. *Pos* is the byte offset of each field relative to the start of the ICMPv6 header. *Len* is the size of each field in bytes.

| Pos | Len | Header | Field | Parsed how? |
| --- | --- | --- | --- | --- |
| 0 | 1 | Reply ICMP | Type | Recorded, must be 1, 2, 3 or 4 |
| 1 | 1 | Reply ICMP | Code | Recorded |
| 2 | 2 | Reply ICMP | Checksum | Checksum check |
| 4 | 4 | Reply ICMP | (Type-specific) | Ignored |
| 8 | 4 | Probe IP | Version / TC / FL | Version must be 6 |
| 12 | 2 | Probe IP | Payload length | Must be at least 40 |
| 14 | 1 | Probe IP | Next header | Must be 58 |
| 15 | 1 | Probe IP | Hop limit | Recorded for distance calculation |
| 16 | 16 | Probe IP | Source address | Must be source address |
| 32 | 16 | Probe IP | Destination addr. | Must be target address |
| 48 | 1 | Probe ICMP | Type | Must be 128 |
| 49 | 1 | Probe ICMP | Code | Must be 0 |
| 50 | 2 | Probe ICMP | Checksum | Ignored |
| 52 | 4 | Probe ICMP | Ident. + Seq. nr. | Ignored |
| 56 | * | Probe Data | Data | Ignored |

## 5.5. Response handling

Two types of responses are expected: ICMPv6 echo replies and ICMPv6 error messages. Echo replies look much the same as the echo request that was sent. (Table 5.1) The responding node only changes the type field to 129 and updates the checksum.

ICMPv6 error messages are much more complex, as they contain the packet that caused the message to be sent as a payload. (See columns one to four of Table 5.2) The ICMPv6 error message header is followed by the IPv6 header, the ICMPv6 header and the data of the probe packet.

Incoming ICMPv6 packets are handled as follows: Firstly, it is assumed that the last 40 bytes of the packet contain the token and the MAC. Packets with an invalid MAC are discarded. Afterwards, a parser attempts to extract certain information from the packet, which is added to the output file on success. On parser failure, `probe.py` prints the packet.

The parser operates as follows: If the type of the ICMPv6 packet is 129 (echo response) and the code is 0, the IPv6 source address is recorded and the parsing is finished. Else, if the type is not 1, 2, 3 or 4 (errors), the parsing fails.

ICMPv6 error messages are parsed as summarized in Table 5.2 and explained in the following: The type and code fields are recorded and the checksum checked. The type-specific data field of the ICMPv6 error message header is ignored. The attached IPv6 packet (which should be the probe packet we sent) is parsed as follows: The version field must be 6, the payload length at least 40 and the next-header type 58 (indicating

ICMPv6). The hop limit is recorded and used to estimate the number of hops *(distance)* on the path from the scanning to the responding host later on.[2] The source address is checked to be the source IP address used by `probe.py` and the destination address must be the target IP address the probe packet was originally sent to. The ICMPv6 payload must have type set to 128 and code set to 0.

To conclude, the following information is recorded for each successfully parsed incoming packet:

**target** The destination IP address the probe packet was sent to

**responder** The source IP address of the response

**type/code** The ICMPv6 type and code fields of the response

**distance** If the response is an ICMPv6 error message: Responder distance estimation (Original hop limit of the probe packet minus hop limit of the probe packet inside the ICMPv6 error message)

## 5.6. Output format

The output file written by `probe.py` is a ZIP archive containing JSON files. The archive file `targets.json` contains the target list as supplied to `probe.py` on startup. Furthermore, the archive contains one directory per range in the target list, each named after its index in the list. Each range directory contains a file named `metadata.json` containing information such as the source IP address used for the scan, scan duration, timestamps of start and end of the scan and the hop limit used.[3] A second file in each directory, `responses.json`, contains the list of responses that were received, as described in the previous section.

## 5.7. Operational and ethical considerations

Internet connectivity for my scanning host was kindly provided by AS206226 — a non-commercial network based in Bremen, Germany. During my scans, AS206226's only transit provider was Hurricane Electric (AS6939), an extremely well connected global carrier.

Due to my decision not to prioritise performance while developing `probe.py`, it was impossible to send more than ten thousand packets per second (approx. 8 Mbit/s), limited by the single-core performance of the scanning host. This packet rate is negligible, even when directed only at a single node. The maximum peak packet rate any CPE had to respond to was 0.2 packets per second.

---

[2]I write 'estimate' because there is an inconsistency between implementations when a *time exceeded* error is generated: Some routers attach the original packet as it was received (with hop limit set to 1), while others attach the packet as it would have been forwarded (with hop limit set to 0).

[3]This is indeed mostly information that is exactly the same for every range that was scanned and it is stored in this way only for historical reasons.

In case someone noticed my scans and got interested, I hosted an informative website on the source IP address of my scans, detailing the nature of my research and providing a way to opt-out of it. I received no opt-out request. Furthermore, AS206226 was not contacted regarding my scans.

To control correct operation of `probe.py`, I configured a Linux host to respond to any IP address inside a /64 prefix and added a sub-prefix to the target list on every scan.

## 5.8. Suffix considerations

In this thesis, I will always run `probe.py` with target suffixes of at least 64-bit length. Consequently, some thought should be put into choosing the last 64 bits of each probed address, as it will be the interface identifier *(IID)* if the first 64 bits are a valid subnet address. (see section 2.3)

Beverly and coauthors propose to choose the IID at random, so that no subnet host with that address exists and the router has to respond with an *address unreachable* ICMP error message. [1, §4.3]

Instead, I propose to use the subnet-router anycast address, which has the last 64 bits set to zero. [7, §2.6.1] Like a random IID, this address will never reach a subnet host. But, unlike a random IID, the router will never even start a neighbor solicitation process to check if a subnet host with the given IID exists. Instead, the router itself is addressed by the probe packet. On typical CPEs and other routers running Linux, this will cause the probe packet to be processed differently by the firewall, since it is no longer a packet aiming to be forwarded by the router (`FORWARD` chain), but a packet aiming to be processed by the IP stack of the router itself (`INPUT` chain).

## 5.9. Performance evaluation

I found that `probe.py` can send and receive at a combined rate of up to 8000 packets per second, before exceeding the single-core performance of a 2017 Intel Xeon Gold processor. This surprisingly low number is the result of both the extremely bad runtime performance of Python code and my questionable decision to use a message authentication code, in about equal shares. On the bright side, this makes it pretty much impossible to accidentally break networks with `probe.py`. Furthermore, I found the probing performance to be sufficient for my scans and felt a lot more limited by the performance of my data analysis scripts.

# 6. Case study: Major German residential broadband networks

To evaluate the applicability of the methods introduced in this thesis to typical residential broadband networks, this chapter will apply them to the three major German IAPs: Deutsche Telekom, Vodafone Kabel Deutschland and 1&1 (United Internet). In total, the three companies have 25 million customers (71% market share). You can find more information about the German residential broadband market in section 2.5.

In the following, I will make use of the aggregated IPv6 hitlists I compiled in chapter 4, my simple probing method for IAP networks (chapter 5), and IPv6 address space visualisation as described in chapter 3.

## 6.1. In preparation: Determining customer prefix length

To enable efficient probing, I first determined the sizes of the prefixes routed to each customer. The smallest useable customer prefix would be a /64, which would allow for just a single LAN at the customers residence. The broadband forum recommends assigning prefixes of size /56. (More on this in section 2.4)

My approach is as follows: First, a prefix under test is probed (as described in chapter 5) at /64 resolution (i.e. one probe packet is sent into each /64 in the prefix). Then, for every responder (i.e. node which responded to a probe packet), I define the *prefix of responsibility* (PR) as the most specific prefix containing all probed addresses that elicited a reply of the given responder. In the special case of only a single reply, the PR is the encompassing /64, instead of a /128, because it would not be sensible for an IAP to route a prefix longer than /64 to a customer. Finally, all responders with a PR equal to the prefix under test (i.e. responding all over the place) are assumed to be the routers connecting the CPEs to the IAP's network and discarded.

I conducted my measurement as follows: For each of the three ISP networks considered in this chapter, 32 prefixes were selected at random from hitlist48. (see section 4.3) Again at random, a single /52 prefix was selected from each /48 prefix. (implemented in `gen-targets-plen.py`, see section A.1) The resulting 96 prefixes were probed at /64 resolution; 393 thousand probe packets were sent within 20 minutes on 2020-08-11. For all responders, the prefix of responsibility was determined and their lengths were aggregated for each ISP. (implemented in `plen.py`, see section A.1)

Figure 6.1 shows the resulting histograms. It is visible that both Deutsche Telekom and 1&1 are routing /56 prefixes to their customers, as recommended. Vodafone uses /62 and /64 prefixes.

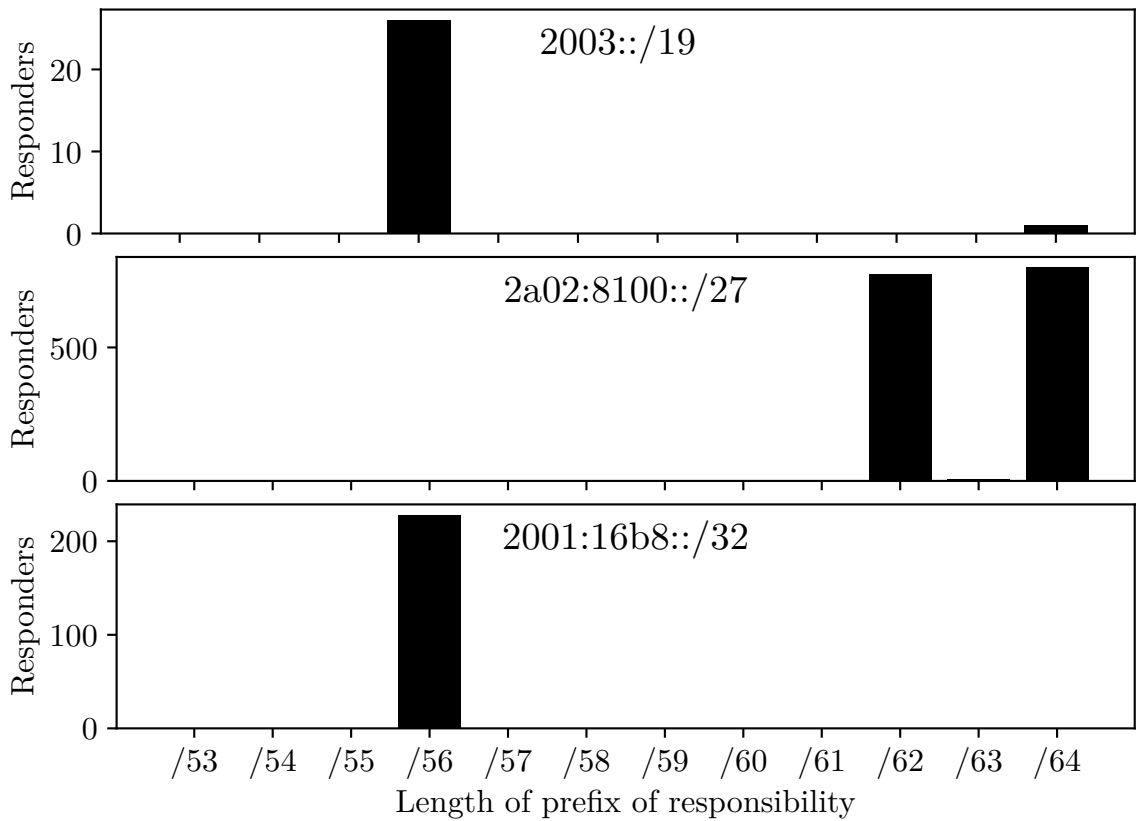

Figure 6.1.: Estimated PR length distributions of three IAP networks. Note the different y axis scales.

## 6.2. Deutsche Telekom

Deutsche Telekom's largest IPv6 address space allocation is 2003::/19. This is also the allocation containing the most active addresses, as I demonstrated in chapter 4, Table 4.1. Figure 6.2 shows the distribution of hitlist56 entries over the entire allocation. It is apparent that most of the space is currently unused and that 2003:c0::/26 is a particularly active region. Indeed, 96 percent of all source addresses from 2003::/19 seen at my NTP server are in that region.

Figure 6.3 zooms into 2003:c0::/26 for a more detailed look, which reveals about a thousand similar-looking /37 prefixes. Figure 6.4 provides an even more detailed look. Most of the /37s have a customer prefix pool in the last /40; some of them also have one in the second half of the fourth /40.

To prepare probing for customer prefixes and CPEs, I created a target definition from all hitlist48 entries inside 2003:c0::/26. (`gen-targets-dtag.py`) On 2020-08-22, `probe.py` sent 38 million probe packets over two hours.

Over the entire runtime, it reported a few response parsing errors. They were caused by response packets that failed the check of the destination address in the original probe IPv6 header. (see chapter 5: position 32 in Table 5.2 and section 5.5) Apparently, my probe packets had been prefix-translated on the path to the responding CPE. The translated addresses were all inside 2003:6::/32. This is quite exciting, because it is the first time that I have seen IPv6 network address translation (NAT) in the wild, but I did not investigate further due to the scope of my thesis.

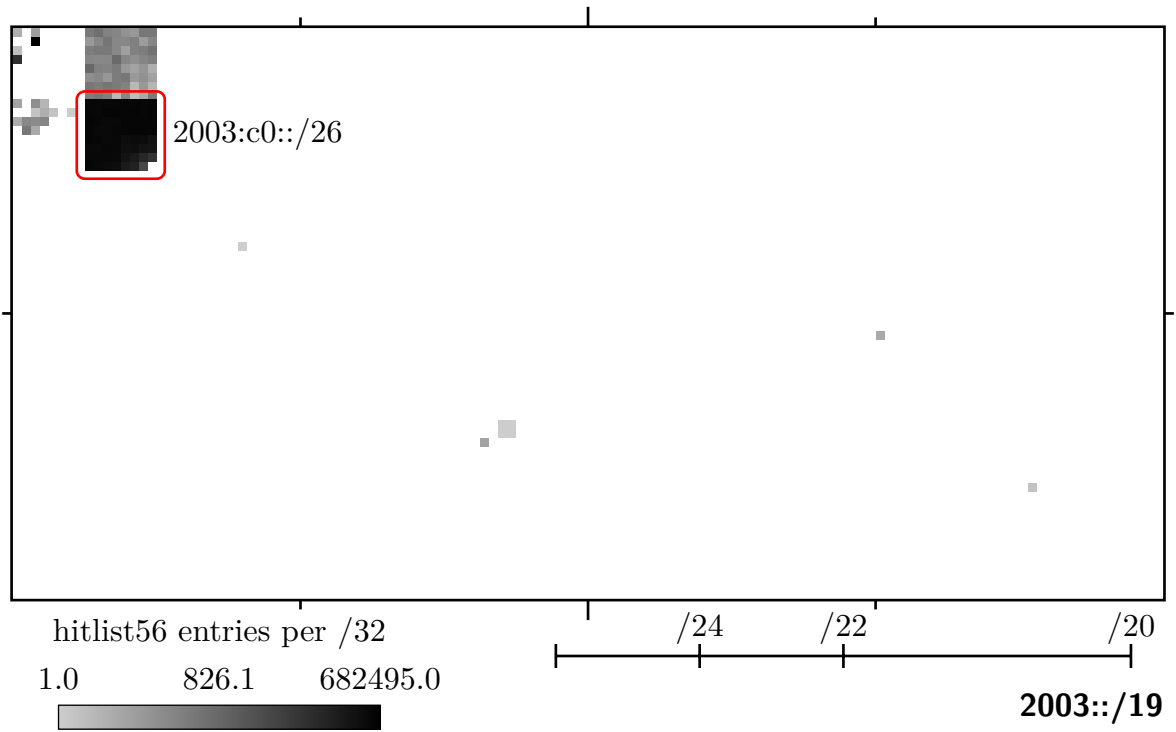

Figure 6.2.: 2003::/19 (Deutsche Telekom): Number of hitlist56 entries per /32. Attention: Logarithmic scale!

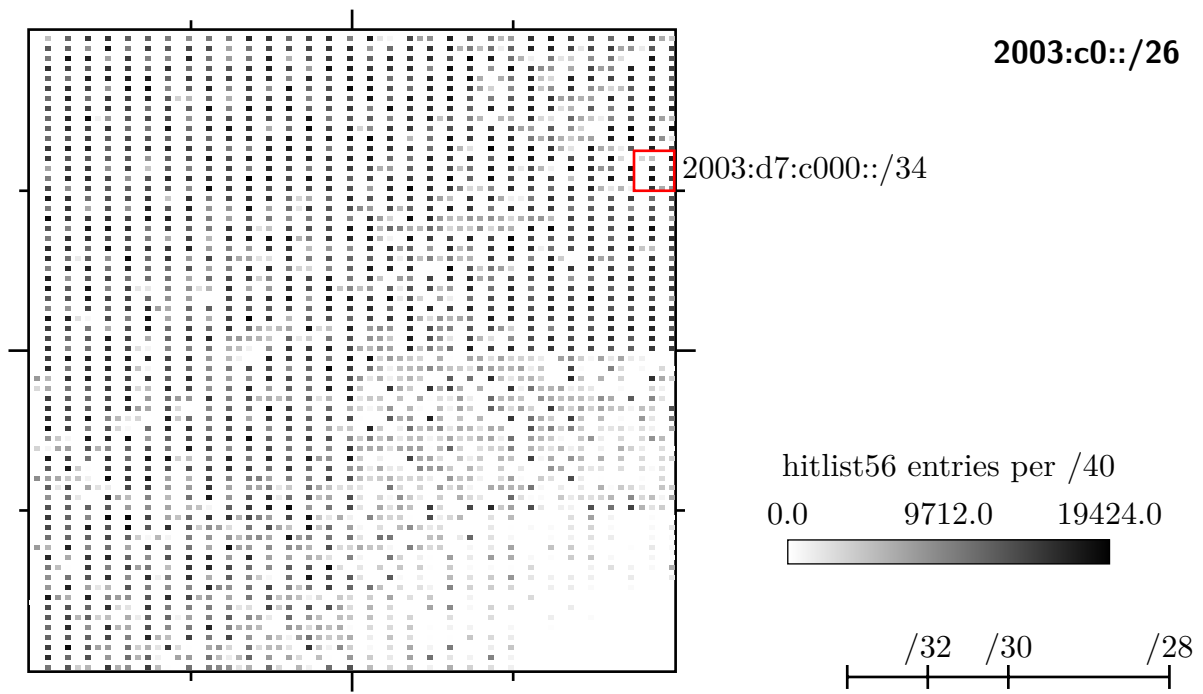

Figure 6.3.: 2003:c0::/26 (Deutsche Telekom): Number of hitlist56 entries per /40.

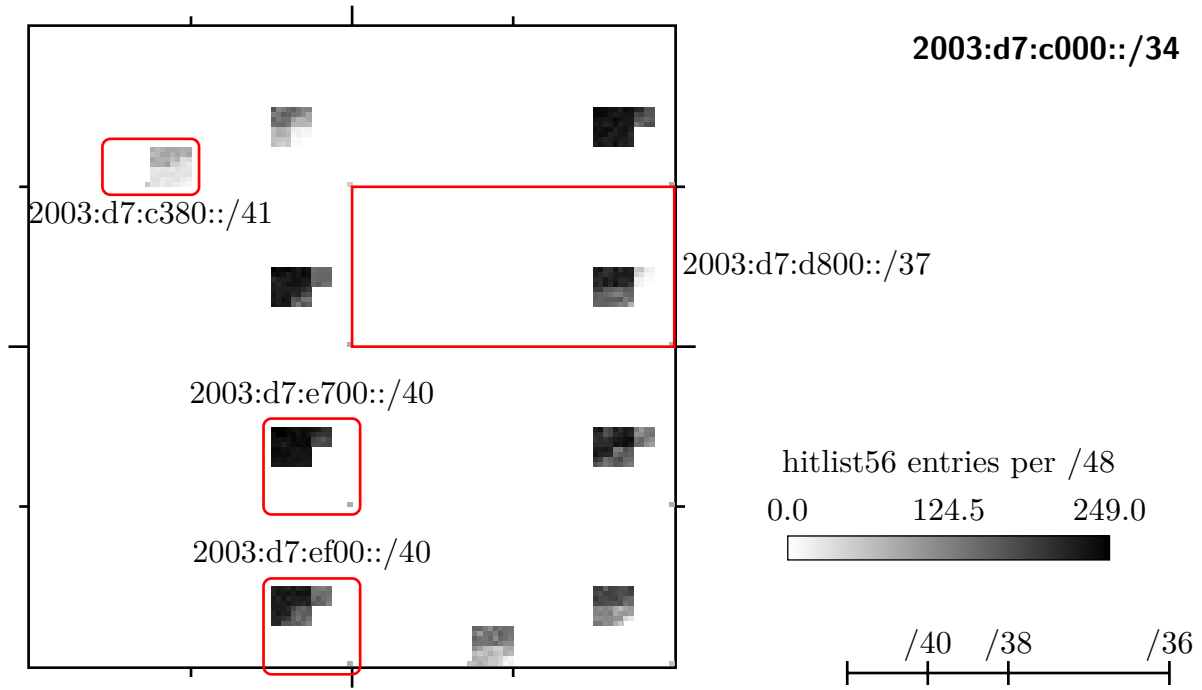

Figure 6.4.: 2003:d7:c000::/34 (Deutsche Telekom): Number of hitlist56 entries per /48.

`probe.py` found 4 million customer prefixes and their corresponding CPEs. Almost all CPEs responded using addresses derived from a modified EUI-48. According to the EUI-48s, 94 percent of the CPEs were manufactured by AVM and another 6 percent by Huawei.

## 6.3. Vodafone Kabel Deutschland

According to chapter 4, Table 4.1, the most active IPv6 address space allocation of Vodafone Kabel Deutschland is 2a02:8100::/27. Figure 6.5 shows the distribution of hitlist64 entries over that allocation. As you can see, only the first half of the address space is used at the moment. Most active spots seem to be concentrated in 2a02:8108::/29, except for 2a02:8106::/38, which contains fixed IP address space for business clients. [10] To show the layout of the customer prefix pools in more detail, Figure 6.6 zooms in on 2a02:810d:9000::/36.

To prepare probing for customer prefixes and CPEs, I created a target definition from all hitlist52 entries inside 2a02:8108::/29. (`gen-targets-vkd.py`) On 2020-08-21, `probe.py` sent 24 million probe packets over two hours, while reporting no parser errors.

`probe.py` found 1.9 million customer prefixes and their corresponding CPEs. Almost no CPEs responded using addresses derived from a modified EUI-48. Based on all responses I received, I reapplied the customer prefix length determination approach from section 6.1 to get a more precise measurement of the prefix length distribution. Figure 6.7 shows the result.

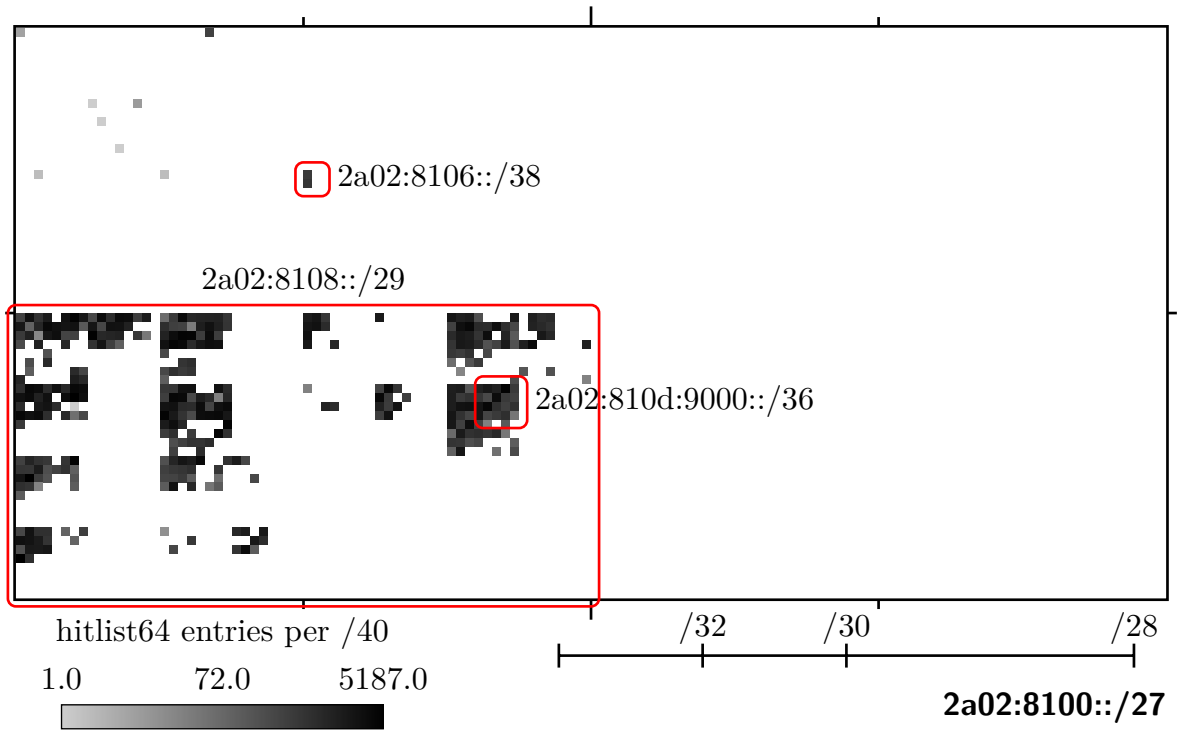

Figure 6.5.: 2a02:8100::/27 (Vodafone Kabel Deutschland): Number of hitlist64 entries per /32. Attention: Logarithmic scale!

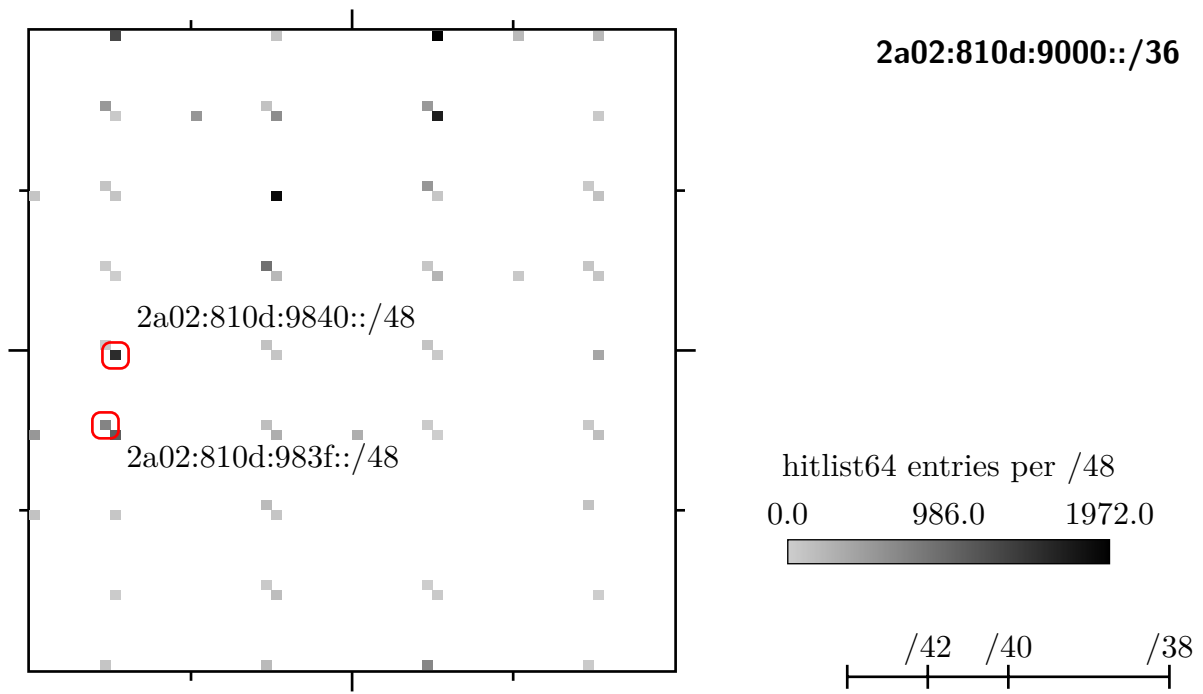

Figure 6.6.: 2a02:810d:9000::/36 (Vodafone Kabel Deutschland): Number of hitlist64 entries per /48.

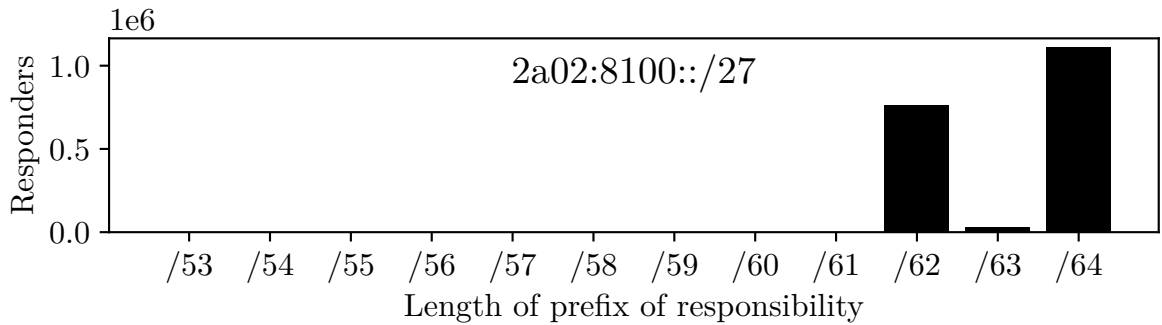

Figure 6.7.: Measured PR length distributions of all probed parts of the Vodafone Kabel Deutschland network.

### 6.3.1. Routing loops

While probing the networks of Deutsche Telekom and 1&1, I never received any ICMPv6 error messages from the last hop previous to the CPEs, even though they are supposed to send (rate-limited) 'no route to destination' errors in response to probe packets that are targeted at currently unused address space. I am confident that this is because the routers were configured not to send these errors. In contrast, while probing the Vodafone Kabel Deutschland network, I received very many responses from their routing infrastructure. (84% of all responses received)

Alarmingly, those responses are not in fact 'no route to destination' errors sent by the last hop previous to the CPEs, but 'hop limit exceeded in transit' errors, which are indicative of routing loops: Let us assume — for the sake of explanation — that the packets sent to the customer prefixes inside 2001:db8:42::/48 are supposed to be routed via router A, to router B, and finally to the CPE in question. Furthermore, packets from the customer prefixes to the Internet are supposed to be routed via router B to router A, where dynamic routing via BGP begins. The network operator can implement this routing policy by configuring router A with a static route to 2001:db8:42::/48 via B and configuring router B with a static default route (::/0, matches all destinations) via A. Additionally, a route to each customer prefix via the matching CPE will be configured on router B.

This routing setup works fine, unless a packet needs to be routed to an address inside 2001:db8:42::/48 which is not inside a customer prefix. Router A will forward it to router B, adhering to the static route to 2001:db8:42::/48. Router B will forward it back to router A, because none of the customer prefix routes match and the default route is used as a last resort. In consequence, the packet is forwarded in a loop between A and B, until the *hop limit* field in the IPv6 header reaches 0, the packet is discarded and an ICMPv6 'hop limit exceeded in transit' error is sent.

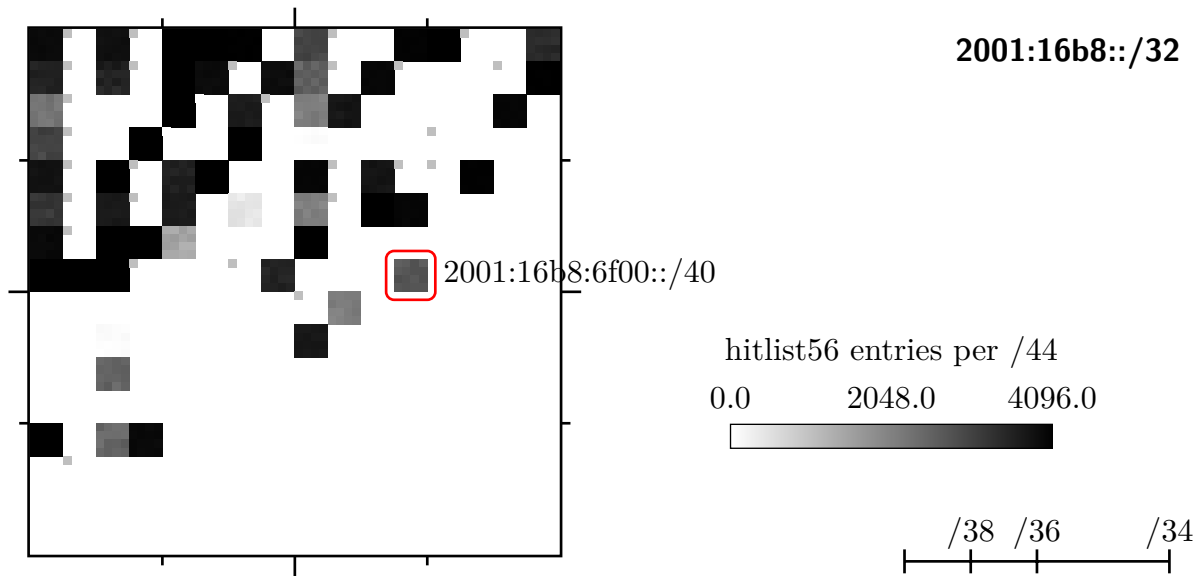

Figure 6.8.: 2001:16b8::/32 (1&1): Number of hitlist56 entries per /44.

Routing loops make networks vulnerable to denial-of-service attacks. In this case, the loops involve just two routers each, so a single packet sent by an attacker with the maximum hop limit of 255 will traverse the link between the two routers more than a hundred times in each direction before it is discarded. In consequence, an attacker sending packets with 100 Mbit/s is able to overload a 10 Gbit/s link.

## 6.4. 1&1 (United Internet)

As indicated by Table 4.1, 1&1 uses the address space allocation 2001:16b8::/32 for their residential customers. The distribution of hitlist56 entries over the allocation is shown in Figure 6.8. It is clearly visible that 1&1 has partitioned this space into /40s, which are each used either for customer prefixes or for WAN links, in which case only the first /46 is in use.[1] Figure 6.9 shows one of the customer prefix pools in more detail. It is apparent that 1&1 assigns customer prefixes uniformly over the entire /40.

To prepare probing for customer prefixes and CPEs, I created a target definition from all hitlist48 entries inside 2001:16b8::/32. (`gen-targets-1u1.py`) On 2020-08-13, `probe.py` sent 3.8 million probe packets over twelve minutes, while reporting no parser errors.

`probe.py` found 1.6 million customer prefixes and their corresponding CPEs. Almost all CPEs responded using addresses derived from a modified EUI-48. According to the EUI-48s, 100 percent of the CPEs were manufactured by AVM. It is remarkable that 1&1 operates some customer prefix pools with an utilization of at least 99 percent, meaning that almost every probe packet sent to the /40s in question caused a CPE to respond.

---

[1]This can be confirmed through whois. The descriptions of route6 objects for WAN links contain 'WAN-TN' ('transfer network'); those for customer prefix pools contain 'LAN-DUAL' or 'LAN-DSLITE', probably indicating whether the customers are dual-stacked (CPE still gets a public IPv4 address) or use dual-stack lite (DS-Lite), a transition technology where the CPE is assigned a private IPv4 address and a carrier-grade NAT is used to reduce IPv4 address usage.

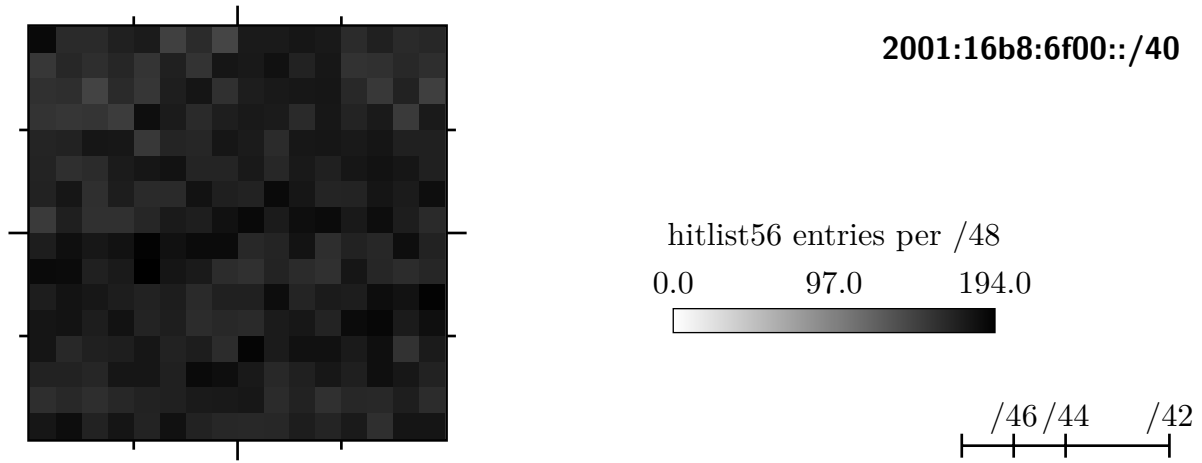

Figure 6.9.: 2001:16b8:6f00::/40 (1&1): Number of hitlist56 entries per /48.

## 6.5. Evaluation

I sent 0.4 million probe packets in a preparatory probing run and 66 million in total in the three main probing runs, which found 7.6 million customer prefixes and CPEs. On average, `probe.py` found 0.11 CPEs per packet sent. A summary of the performance of the three main probing runs is provided in Table 6.1.

Table 6.1.: Key data regarding the three main probing runs.

|  | D. Telekom 2003:c0::/26 | Vodafone K. D. 2a02:8108::/29 | 1&1 2001:16b8::/32 | Total |
|---|---|---|---|---|
| Probe packets sent | 38 166 016 | 24 477 696 | 3 801 088 | 66 444 800 |
| Responses received | 4 045 927 | 12 169 939 | 1 630 320 | 17 846 186 |
| Unique responders | 4 044 355 | 1 900 168 | 1 630 311 | 7 574 834 |
| with EUI-64 | 4 036 587 | 385 | 1 630 009 | 5 666 981 |

# 7. Conclusion and outlook

> IPv4 ist derzeit vollkommen ausreichend für Privatkunden.
>
> *(Kundendienst von EWE)*

In this thesis, I contributed the following: Firstly, I adopted a forgotten method of visualizing the IPv4 address space for use with IPv6 and compared this method to the currently popular method of address space visualization. Secondly, I presented the NTP Pool Project as a publicly available opportunity for passively compiling IPv6 hitlists containing mostly client hosts, 'Smart Home' devices and CPEs. To the best of my knowledge, this is the first public hitlist source aimed at these types of devices described in the literature. Thirdly, I described a scanning technique for Internet access provider *(IAP)* networks, which differ from other networks in their regular structure and high router density, and developed a tool implementing this technique.

Finally, I tested my contributions by applying them to the three major German residential broadband networks, which have 25 million customers (71% market share) in total. I was able to visualize the structure of the three networks based on the IPv6 hitlists I created. I probed 67 million IP addresses and collected responses from 7.5 million CPEs. This accounts for only 30% of the customer base, so it is likely that my method of probing elicited responses from only a portion of reachable CPEs.

75% of the responding CPEs are using IP addresses derived from their Ethernet MAC addresses. This type of address enables telemarketers and others to recognize users of a certain CPE (i.e. a household) over the entire lifetime of the CPE.

## 7.1. Comparison with Rye and Beverly

While writing my thesis, long after I finished the literature research, I found a 2020 paper by Rye and Beverly, which is quite relevant and complements my work. [29] I will summarize their work and compare our results in the following.

Their paper considers the discovery of the IPv6 network *periphery*, which they define to consist of all last hop routers connecting hosts to the Internet. They introduce CPEs in residential networks as a prime example of the periphery. Thinking of their problem as one of topology discovery, they apply yarrp6 to probe iteratively refined lists of target addresses. Somewhat surprisingly, their algorithm only considers last hop information from the yarrp6 output, so the bulk of the probe packets were sent unneccessarily.

To seed their search, they use and compare two approaches: Firstly, a *BGP-informed* seed consisting of traceroutes conducted by CAIDA in 2014/2015 to every /48 inside of a questionable subset of BGP announcements.[1] Secondly, a *hitlist-informed* seed consisting either of traceroutes they conducted to entries of the regularly updated IPv6 hitlist published by Gasser and coauthors [9] or traceroutes conducted in a prior effort by Beverly and coauthors.[2] [1]

In total, Rye and Beverly used yarrp6 to probe at least 1022 million IP addresses. Based on the information they provided about how they used yarrp6, I estimate that they have sent at least $11 \times 10^9$ probe packets. They discovered 64 million periphery router addresses over the course of two months, but realized that this number overestimates the number of discovered periphery routers, because of *prefix rotation*: Some IAPs change the prefix assigned to each customer on a regular basis, for example daily. (They mention AS 8881 (1&1) as an example of daily rotation.)

According to figure 5 of their paper, using their hitlist-informed seed, they discovered 4 and 1 million CPEs in the networks of Deutsche Telekom and 1&1, respectively, and 9 million CPEs in the network of 1&1 using their BGP-informed seed. Considering their prefix rotation issue, it appears that they found as many CPEs as I did.

To conclude, Rye and Beverly's approach sent a huge amount of packets, most of them unneccessarily. Consequently, their scans took a long time and suffered from artifacts due to prefix rotation as a result. On average, they found at most 0.006 periphery routers per packet sent, while I found 0.1 routers per packet sent with a simpler probing method.

This comparison, however, is not entirely fair, because I scanned only networks I observed to contain client hosts during my NTP Pool hitlist compilation process, which took over half a year. I argue that, firstly, I could have got the same results with a hitlist compiled over a much shorter timeframe, and secondly, as Rye and Beverly have shown, it is possible to find customer networks without this type of hitlist, purely based on iteratively refined sampling of the IPv6 address space, as announced through BGP.

## 7.2. Outlook

To pick up on the last section, it would be interesting to look into purely BGP-informed periphery discovery, as tried by Rye and Beverly, again, but without making use of CAIDAs bad dataset and by combining my simple probing method with Rye and Beverly's iterative refining of the target set. In addition, my approach for responder distance estimation could be used to guide iterative exploration, on the assumption that responses from increasingly more distant routers indicate the use of subnetting.

A future study should compare different kinds of probe packets by their effectiveness in eliciting responses from CPEs:

- Subnet-router anycast address or last 64 bits choosen at random

---

[1] Besides being outdated, this dataset does, by design, not contain any traceroutes to networks larger than a /32, which probably excludes every single IAP network, unless the IAP happens to also advertise /32 more-specifics through BGP for traffic engineering. This is the reason why Rye and Beverly found no periphery in AS 3320 (Deutsche Telekom) when using the BGP-informed seed.

[2] The paper makes it sound as though these two sources were the same, but Beverly and coauthors explicitly state that they did not build on the hitlist by Gasser and coauthors, so I am at a loss here.

- Connection attempts to typical services running on CPEs: SIP, HTTP(S), DNS, NTP, UPNP, IPP, CWMP, FTP, SMB, SSH, Telnet

- Connection attempts to typical rejected ports: Ident (113), NetBIOS (137), SMTP

- Fragmented packets

To complement this study, it would be interesting to collect firewall rulesets from typical CPEs to gather inspiration for more effective probing. This is quite a lot of effort, though, because CPE vendors do not typically intend to provide shell access to the user, so one either has to break into a running CPE or make use of reverse-engineering methods. (i.e. firmware analysis, physically accessing the flash memory, etc.)

In addition, it would be interesting to search for customer networks without an effective firewall and try to enumerate the hosts inside them. This comprises guessing the IID (last 64 bits of the IP address) of the hosts, which is obviously infeasable if they were generated at random as semantically opaque or through privacy extensions,[3] but might be possible for addresses assigned through stateful DHCPv6 or even those derived from Ethernet MAC addresses. The main hurdle for a benign study design is to probe IIDs without exhausting the CPE's neighbor discovery capacity, so that functioning of the CPE is not impacted.

---

[3]But see Ullrich and Weippl [34]

# A. Supplementary material

This chapter lists parts of this thesis that are not published as part of this document due to their nature, length constraints or legal reasons.

## A.1. Source code

All my source code is available here: `https://gitlab.informatik.uni-bremen.de/tbruns/ma-code`

Due to time constraints, I did not decide on a license or add a README file. Feel free to eMail me with any questions.

## A.2. Response datasets

Due to time constraints, I did not publish this dataset, but will do so on request.

## A.3. NTP datasets

Due to time constraints, I did not publish this dataset, but will do so on request.

# List of Figures

# List of Tables